\documentclass[11pt,letter]{article}

\usepackage{hyperref}

\usepackage[english]{babel}
\usepackage[latin1]{inputenc}
\usepackage[T1]{fontenc}

\usepackage{amsmath,amssymb}
\usepackage[mathscr]{euscript}
\usepackage{graphicx}
\usepackage{amssymb}

\usepackage{fullpage}

\usepackage{theorem}
\newtheorem{theorem}{Theorem}

\newtheorem{lemma}[theorem]{Lemma}

\newtheorem{fact}[theorem]{Fact}

\theorembodyfont{\upshape}

\renewcommand{\leq}{\leqslant}
\renewcommand{\geq}{\geqslant}

\newenvironment{proof}{\paragraph{Proof.}}{\hfill$\Box$}
\newenvironment{proofOf}[1]{\paragraph{Proof of {#1}.}}{\hfill$\Box$}

\newcommand{\algonom}[1]{{\textsc{\bfseries{#1}}}\/}
\newcommand{\EE}{{\algonom{B-EquiSet}}}
\newcommand{\LWF}{{\algonom{LWF}}}
\newcommand{\EEs}{{\algonom{B-EquiSet$_s$}}}
\newcommand{\EEDF}{{\algonom{B-EquiSet-Edf}}}

\newcommand{\EA}{{\algonom{Equi$\circ$A}}}
\newcommand{\EsA}{{\algonom{Equi$_s\circ$A}}}
\newcommand{\ESA}[1]{{\algonom{Equi$_{\ensuremath{#1}}\circ$A}}}
\newcommand{\EQ}{{\algonom{Equi}}}

\newcommand{\MINIDX}{{\algonom{MinIdx}}}
\title{Pull-Based Data Broadcast with Dependencies: \\ Be Fair to Users, not to Items}
\author{\large Julien Robert and Nicolas Schabanel\\[3mm]
\normalsize École normale supérieure de Lyon\\ {\small \raisebox{-.25em}{\makebox[0em][r]{\includegraphics[height=1.1em]{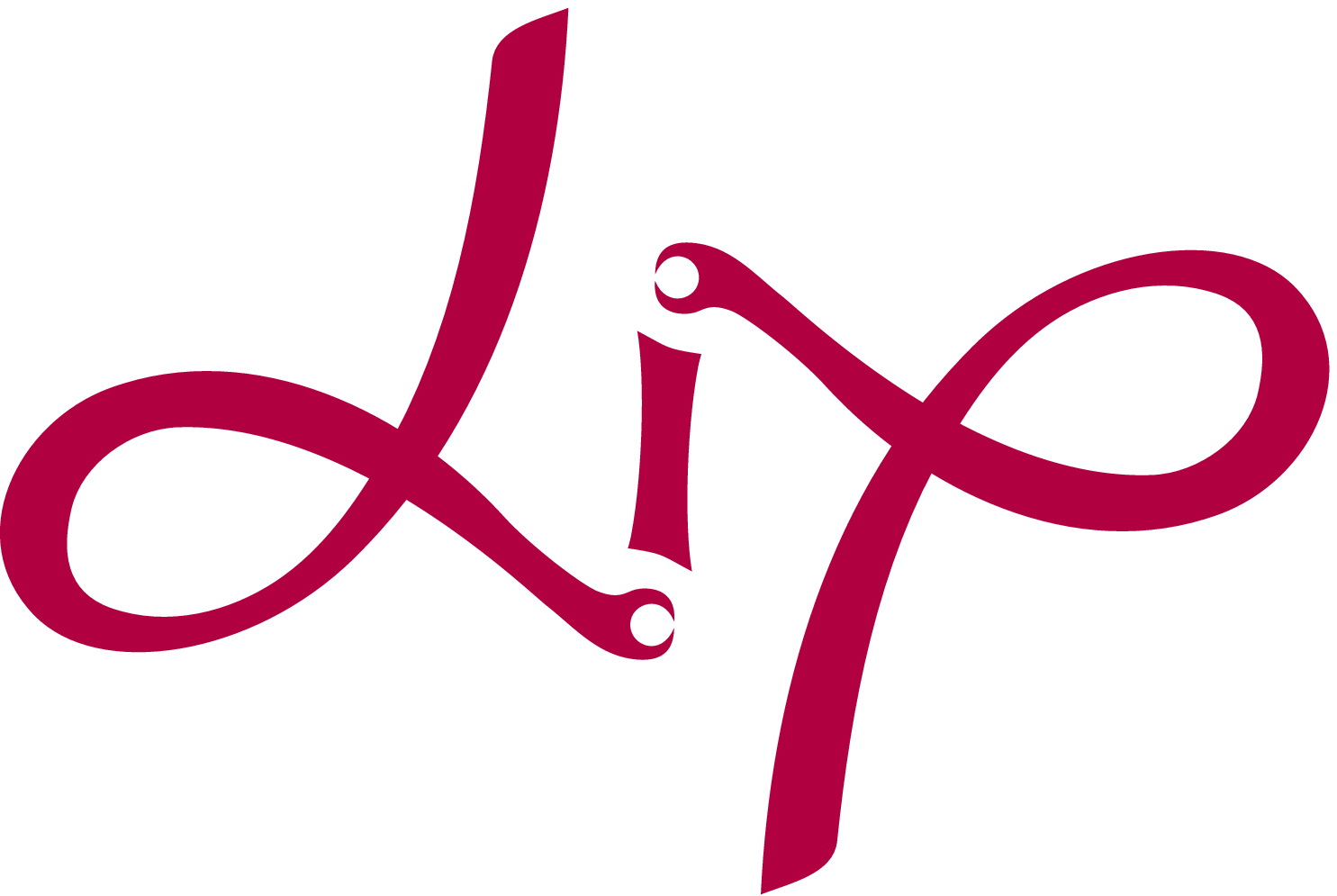}}} Laboratoire de l'informatique du parallélisme}\\{\small UMR CNRS ENS-LYON INRIA UCBL n°5668}\\ \normalsize 46 allée d'Italie, 69364 Lyon Cedex 07, France\\[3mm] \normalsize $\texttt{http://perso.ens-lyon.fr/}\{\texttt{julien.robert}, 
\texttt{nicolas.schabanel}\}$}

\newcommand{\ie}{\textit{i.e.}\/}
\newcommand{\eg}{\textit{e.g.}\/}

\newcommand{\FT}{\ensuremath{\operatorname{B-FlowTime}}}
\newcommand{\JFT}{\ensuremath{\operatorname{J-FlowTime}}}
\newcommand{\BFT}{\ensuremath{\operatorname{\mathscr{B}-FlowTime}}}

\newcommand{\Sc}[1]{\ensuremath{\mathscr{#1}}}

\newcommand{\iset}{{\Sc{I}}}
\newcommand{\bset}{{\Sc{S}}}
\newcommand{\Bset}{{\Sc{B}}}
\newcommand{\jset}{{\Sc{J}}}
\newcommand{\bsch}{{\Sc{E}}}
\newcommand{\osch}{{\Sc{O}}}
\newcommand{\bopt}{{\ensuremath{\operatorname{BOPT}}}}
\newcommand{\jopt}{{\ensuremath{\operatorname{JOPT}}}}
\newcommand{\jsch}{{\ensuremath{\operatorname{\mathfrak{S}}}}}
\newcommand{\Bopt}{{\ensuremath{\operatorname{\mathscr{B}OPT}}}}
\newcommand{\Utwo}{{\ensuremath{\operatorname{\Upsilon\!_2}}}}
\newcommand{\CC}{{\ensuremath{\mathcal{C}}}}
\begin{document}
\maketitle

\thispagestyle{empty}

\begin{abstract}
Broadcasting is known to be an efficient means of disseminating data in wireless communication environments (such as  Satellite, mobile phone networks,...). It has been recently observed that the average service time of broadcast systems can be considerably
 improved by taking into consideration existing correlations between requests. We study a pull-based data broadcast system where users request possibly overlapping sets of items; a request is served when all its requested items are downloaded. We aim at minimizing the average user perceived latency, \ie\/ the average flow time of the requests. We first show that any algorithm that ignores the dependencies can yield arbitrary bad performances with respect to the optimum even if it is given arbitrary extra resources. We then design a $(4+\epsilon)$-speed $O(1+1/\epsilon^2)$-competitive algorithm for this setting that consists in 1) splitting evenly the bandwidth among each requested set and in 2)  broadcasting arbitrarily the items still missing in each set into the bandwidth the set has received. Our algorithm presents several interesting features: it is simple to implement, non-clairvoyant,  fair to users so that no user may starve for a long period of time, and guarantees good performances in presence of correlations between user requests (without any change in the broadcast protocol). We also present a $ (4+\epsilon)$-speed $O(1+1/\epsilon^3)$-competitive algorithm which broadcasts at most one item at any given time and preempts each item broadcast at most once on average. As a side result of our analysis, we design a competitive algorithm for a particular setting of non-clairvoyant job scheduling with dependencies, which might be of independent interest.
 
\paragraph{Keywords:} Multicast scheduling, Pull-based broadcast, Correlation-based, Non-clairvoyant scheduling, Resource augmentation.
\end{abstract}
\vspace*{1cm}
\begin{center}
\itshape
Omitted proofs, lemmas, notes and figures\\ may be found in appendix.%
\footnote{This work is supported by the CNRS Grant \raisebox{-.25em}{{\includegraphics[height=.95em]{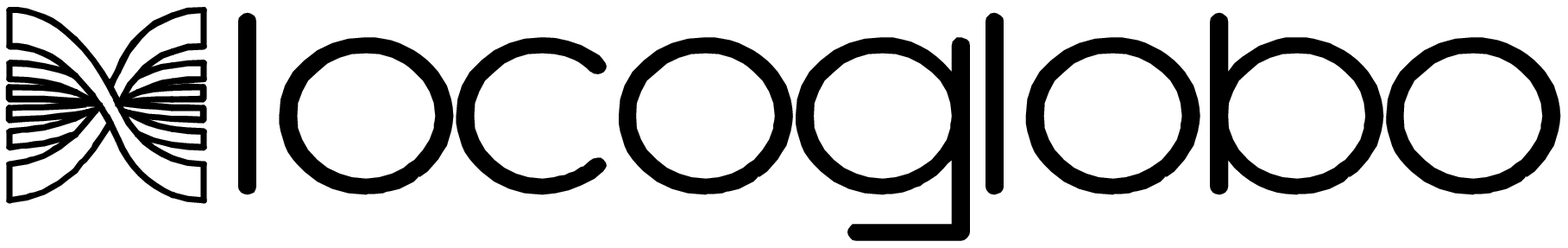}}}.}
\end{center}

\pagebreak

\setcounter{page}{1}

\section{Introduction}

\paragraph{Motivations.}
Broadcasting is known to be an efficient means of disseminating data in wireless communication environments (such as  Satellite, mobile phone networks,...). It has been recently observed  in \cite{HuangChen2004,HuangChen2003,CaiLinChen2005}  that the average service time of broadcast systems can be considerably
 improved by taking into consideration existing correlations between requests.  Most of the theoretical research on data broadcasting was conduct until very recently under the assumption that user requests are for a single item at a time and are independent of each other. However, users usually request several items at a time which are, to a large extent, correlated. A typical example is a web server: users request web pages that are composed of a lot of shared components such as logos, style sheets, title bar, news headers,..., and all these components have to be downloaded together when any individual page is requested. Note that some of these components, \eg\/ news header, may constantly vary over time (size and/or content).\\[-2.5em]

\paragraph{Pull-based data broadcast with dependencies.}
We study a pull-based data broadcast system where users request possibly overlapping sets of items. We aim at minimizing the average user perceived latency, \ie\/ the \emph{average flow time} of the requests, where the flow time of a request is defined as the time elapsed between its arrival and the end of the download of the last requested item. We assume that user  cannot start downloading an item in the middle of its broadcast. When the broadcast of an item starts, all the outstanding requests asking for this item can start downloading it. Several items may be downloaded simultaneously. We consider the \emph{online} setting where the scheduler is \emph{non-clairvoyant} and discovers each request at the time of its arrival; furthermore, the scheduler does not even know the lengths of the requested items and is aware of the completion of a broadcast only at the time of its completion. Items are however labeled with a unique ID to allow their retrieval. Note that this are the typical requirements of a real life systems where items may vary over time.\\[-2.5em]

\paragraph{Background.}
It is well known that preemption is required in such systems in order to achieve reasonable performances. Furthermore, \cite{EdmondsPruhs2003} proved that even without dependencies, no algorithm can guarantee a flow time less than $\Omega(\sqrt n)$ times the optimal. The traditional approach in online algorithms consists then in penalizing the optimum by increasing the bandwidth given to the algorithm so that its performances can be compared to the optimum. This technique is known as \emph{resource augmentation} and provides interesting insights on the relative performances of different algorithms that could not be compared directly to the optimum cost. In our case, we give to our algorithm a bandwidth~$s>1$ and show that it achieves a flow time less than a constant times the optimum cost with a bandwidth~$1$. Formally, an algorithm is  \emph{$s$-speed $c$-competitive} if when given a bandwidth~$s$, its flow time is at most at a factor $c$ of the optimum flow time with bandwidth~$1$. 

To our knowledge the only positive results  \cite{EdmondsPruhs2003,EdmondsPruhs2005} in the online setting assume that the requests are independent and ask for one single item. The authors show that without dependencies the algorithms \EQ\/ and \LWF\/ are competitive. \EQ\/ which splits evenly the bandwidth among the alive requested items, is $(4+\epsilon)$-speed $(2+8/\epsilon)$-competitive, and \LWF, which broadcasts the item where the aggregate waiting times of the outstanding requests for that item is maximized, is $6$-speed $O(1)$-competitive (where the bound proved on the competitive ratio is $O(1)=6,\!000,\!000$). In the \emph{offline} setting, where the requests and their arrival times are known at time $t=0$, the problem is already NP-hard but better bounds can be obtained using linear programming \cite{KalyanasundaramPruhsVelauthapillai2000,ErlebachHall2002,GandhiKhullerKimWan2002,BansalCharikarKhannaNaor2005ondemand,BansalCoppersmithSviridenko2006}; the latest result,  \cite{BansalCoppersmithSviridenko2006} to our knowledge, is a $O(\log^2(T+n)/\log\log(T+n))$-approximation where $n$ is the number of requests and $T$ the arrival time of the last request. To our knowledge, our results are the first provably efficient algorithms to deal with dependencies in the online setting.

Concerning the push-based variant of the problem, where the requests arrival times follow some Poisson process and the requested sets are identically distributed according to a fixed distribution, constant factor approximations exist in presence of dependencies \cite{BarnoyShilo2000,BarnoyNaorSchieber2003,DeySchabanel2006}. The latest result, \cite{DeySchabanel2006}, obtains a $4$-approximation if the requested sets are drawn according to an arbitrary fixed distribution over a finite number of subsets of items.\\[-2.5em]

\paragraph{Our contribution.}
We first show that the performances of any algorithm that ignores the dependencies can be arbitrarily far from the optimal cost even if it is given \emph{arbitrary} extra resources. We then design a $(4+\epsilon)$-speed $O(1+1/\epsilon^2)$-competitive algorithm \EE\/ for the non-clairvoyant data broadcast problem with dependencies. \EE\/ consists in 1) splitting evenly the bandwidth among each requested set and in 2) broadcasting arbitrarily the items still missing  in each set into the bandwidth the set has received. The spirit of the algorithm is that \emph{one should favor the users over the items} in the sense that it splits the bandwidth evenly among the outstanding requested sets and arbitrarily among the outstanding items within each requested set. Our algorithm presents several interesting features: it is simple to implement, non-clairvoyant,  fair to users so that no user may starve for a long period of time, and improves performances in presence of correlations between user requests (without any change in the broadcast protocol). Presicely, we prove that:

\begin{theorem}[Main result] \label{thm:EE}
For all $\delta>0$ and $\epsilon>0$, {\EE} is a ${(1+\delta)(4+\epsilon)}$-speed ${(2+8/\epsilon)(1+1/\delta)}$-competitive algorithm for the online data broadcast problem with dependencies.
\end{theorem}

One could object that  \EE\/ is unrealistic since it can split the bandwidth arbitrarily. But using the same technic as in~\cite{EdmondsPruhs2003}, it is easy to modify \EE\/ to obtain an other competitive algorithm $\EEDF$ (described at the end of section~\ref{sec:EEDF}) which, with a slight increase of bandwidth, ensures that at most one item is broadcast at any given time and that each broadcast is preempted at most once on average.  

\begin{theorem}[Bounded preemption] \label{thm:eedf}
For all $\delta>0$ and $\epsilon>0$, {\EEDF} is a ${(1+\delta)^2(4+\epsilon)}$-speed $(2+8/\epsilon)(1+1/\delta)^2$-competitive algorithm for the online data broadcast problem with dependencies, where each broadcast is preempted at most once on average. 
\end{theorem}

Our analysis takes its inspiration in the methods developed in \cite{EdmondsPruhs2003}. In order to extend their analysis to our algorithm, we have also designed a new competitive algorithm \EA\/ for a particular setting of non-clairvoyant job scheduling with dependencies which might be of independent interest (Theorem~\ref{thm:EA}).

\medskip

The next section gives a formal description of the problem and shows that it is required to take dependencies into account to obtain a competitive algorithm. Section~\ref{sec:alg:EE} exposes the algorithm \EE\/ and introduces useful notations. Section~\ref{sec:EA} designs a competitive algorithm \EA\/ for a variant of job scheduling with dependencies that is used in Section~\ref{sec:EE} to analyze the competitiveness of our algorithm \EE\/.

\section{Definitions and notations}

\label{sec:not}
%:\label{sec:not}

\paragraph{The problem.} 
The input consists of:\\[-2em]
\begin{itemize}
\item A set $\iset$ of $n$ \emph{items} $I_1,\ldots,I_n$ each of length $\ell_1,\ldots,\ell_n$\\[-2em]
\item A set $\bset$ of $q$ \emph{requests} for $q$ non-empty sets of items $S_1,\ldots, S_q\subseteq \iset$, with arrival times $a_1,\ldots,a_q$.\\[-2em]
\end{itemize}

\paragraph{Schedule.}
A \emph{$s$-speed schedule} is an allocation of a bandwidth of size $s$ to the items of $\iset$ over the time. Formally, it is described by a function $r : \iset \times [0,\infty) \rightarrow [0,s]$ such that for all time $t$, $\sum_{I \in \iset} r(I,t) \leq s$; $r(I,t)$ represents the \emph{rate} of the broadcast of~$I$ at time~$t$, \ie, the amount of bandwidth allotted to item $I$ at time $t$. An item $I_i$ is broadcast between $t$ and $t'$ if its broadcast starts at time $t$ and if the total bandwidth allotted to $I_i$ between $t$ and $t'$ sums up to $\ell_i$, \ie, if $\int_{t}^{t'} r(I_i,t)\,dt = \ell_i$. We denote by $c(I_i,k)$ the date of the completion of the 
$k$th broadcast of item~$I_i$. Formally,
it is the first date such that $\int_{0}^{c(I_i,k)} r(I_i,t) dt = k\, \ell_i$ (note that $c(I_i,0) = 0$). We denote by $b(I_i,k)$ the date of the beginning of the $k$th broadcast of item~$I_i$, \ie\/ $b(I_i,k) = \inf\{t \geq c(I_i,k-1)\,:\, r(I_i,t) > 0\}$.%
\footnote{Remark that this formalization prevents from broadcasting the same item twice at a given time or from aborting the current broadcast of an item. The first point is not restrictive since if two broadcasts of the same item overlap, one reduces the service time by using the beginning of the bandwidth allotted to the second broadcast to complete earlier the first, and then the end of the first to complete the second on time. The second point is at our strict disadvantage since it does not penalize an optimal schedule that would never start a broadcast to abort it later on.}\\[-2em]

\paragraph{Cost.}
For all time $t$, let $B(I_i,t)$ be the time of the beginning of the first broadcast of item $I_i$ after $t$, \ie\/ $B(I_i,t) = \min\{b(I_i,k):b(I_i,k)\geq t\}$. For all time $t$, $C(I_i,t)$ denotes the time of the end of the first broadcast of item $I_i$ starting after $t$, \ie\/ $C(I_i,t) = \min \{c(I_i,k):b(I_i,k)\geq t\}$. 
The \emph{completion time} $c_j$ of request $S_j$ is the first time such that every item in $S_j$ has been broadcast (or downloaded)  after its arrival time $a_j$, \ie, ${c_j = \max_{I_i\in S_j} C(I_i,a_j)}$.
We aim at minimizing the \emph{average completion time} defined as ${\frac{1}{q} \sum_{S_j \in \bset} (c_j - a_j)}$, or equivalently the \emph{flow time} defined as the sum of the waiting times, \ie\/ $\FT = {\sum_{S_j \in \bset} (c_j - a_j)}$. We denote by $\bopt_s(\bset)$ the flow time of an optimal $s$-speed schedule for a given instance $\bset$.\\[-2em]

\paragraph{$s$-Speed $c$-Competitive Algorithms.}
We consider the online setting of the problem, in which the scheduler gets informed of the existence of each request $S_j$ at time $a_j$ and not before. The scheduler is not even aware of the lengths $(\ell_i)_{I_i\in S_j}$ of the requested items in each set nor of the total number $n$ of available items.               
It is well known (\eg, see \cite{EdmondsPruhs2003}) that in this setting, it is impossible to approximate within a factor $o(\!\sqrt{n})$ the optimum flow time for a given bandwidth~$s$ even if all items have unit length
(independently of any conjecture such as $P=NP$). The traditional approach in online algorithms consists then in penalizing the optimum by increasing the bandwidth given to the algorithm so that its performances can be compared to the optimum. This technique is known as \emph{resource augmentation} and provides interesting insights on the relative performances of different algorithms that could not be compared directly to the optimum cost. In our case, we give to our algorithm a bandwidth~$s>1$ and show that it achieves a flow time less than a constant times the optimum cost with a bandwidth~$1$. Formally, an algorithm is  \emph{$s$-speed $c$-competitive} if when given $s$ times as many resources as the adversary, its cost is no more than  $c$ times the optimum cost. 
In our case the resource is the bandwidth, and we compare the cost~$A_s$ of a scheduler $A$ with a bandwidth~$s$, to the cost $\bopt_1$ of an optimal schedule on a unit bandwidth. (We denote by $A_s$ the cost of an algorithm~$A$ when given a bandwidth $s$.)

We show below that ignoring existing dependencies can lead to arbitrarily bad solutions.

\begin{fact}[Dependencies cannot be ignored]
\label{fac:yao}
%:{fac:yao}
No algorithm $A$ that ignores dependencies is $s$-speed $c$-competitive for any $c < \frac{2}{3s}\sqrt{n}$ if $A$ is deterministic, and for any $c < \frac1{6s} \sqrt n$ if $A$ is randomized.
\end{fact}

\begin{proof}
Consider first a deterministic algorithm $A$ which is given a bandwidth $s$ and consider the instance where $n$ different items are requested at time $t=0$. Since $A$ ignores the dependencies, we set them after the execution of the algorithm $A$: one request asks for the $n-\sqrt n$ items  that have been served the most by $A$ at time $t=(n-\sqrt n)/s$, and $\sqrt{n}$ requests ask for each of the remaining $\sqrt{n}$ items. Then, algorithm $A$ serves each request only after time $t=(n-\sqrt{n})/s$ and its flow time is at least $(\sqrt{n}+1)(n-\sqrt{n})/s \sim n\sqrt n/s$. The optimal solution with bandwidth only~$1$ first broadcasts the items corresponding to the $\sqrt{n}$ unit length requests and then broadcasts the $n-\sqrt{n}$ remaining items; the optimal flow time is then $(n+\sum_{k=1}^{\sqrt{n}} k) \sim \frac32 n$. This shows a gap of $\frac2{3s}\sqrt{n}$ between the optimal cost with bandwidth~$1$ and every deterministic algorithm with bandwidth $s=O(\sqrt{n})$, which ignores the dependencies. 
We extend the result to randomized algorithms thanks to Yao's principle \cite{Yao1977,MotwaniRaghavan1995} (Omitted). 
\end{proof}

\section{The Algorithm \EE}
\label{sec:alg:EE}

\paragraph{Definitions.} 
A request $S_j$ for a subset of items is said to be \emph{alive} at time $t$ if $t\geq a_j$ and if the download of at least one item $I_i\in S_j$ is not yet completed at time $t$, \ie, $t < C(I_i,a_j)$. We say that an item $I_i\in S_j$ whose download is not yet completed (\ie, such that $a_j \leq t < C(I_i,a_j)$) is \emph{alive for $S_j$} at time $t$.\\[-2em]

\paragraph{The {\EE} Algorithm.} 
Consider that we are given a bandwidth $s$. Let $R(t)$ be the set of alive requests at time $t$ during the execution of the algorithm. For all $t$, \EE\/ allocates to each alive request the same amount of bandwidth, $s/|R(t)|$; then, for each alive request $S_j$, it splits \emph{arbitrarily} the $s/|R(t)|$ bandwidth allotted to $S_j$ among its alive items. Precisely, it allocates to each item $I_i$ alive for $S_j$ at time~$t$, an \emph{arbitrary} amount of bandwidth, $r_{j,i}(t) \geq 0$, such that $\sum_{\text{$I_i$ alive for $S_j$}} r_{j,i}(t) = s/|R(t)|$. \EE\/ then broadcasts at time $t$ each item $I_i$ at a rate $r_i(t) = \sum_{S_j\in R(t)\,:\,\text{$I_i$ is alive for $S_j$ at time $t$}} r_{j,i}(t)$. 

\begin{figure}[hb]
\noindent
\begin{tabular}[c]{p{5cm}@{}c}
\footnotesize
\textbf{The instance} consists of three items $A,B,C$ of length $1.5$ and four requests $S_1=\{A,B,C\}$ (in red), $S_2=\{A\}$ (in green), $S_3=\{B\}$ (in blue), and $S_4=\{C\}$ (in yellow) with arrival times $a_1=0$, $a_2=1$, $a_3=2$, and $a_4=3$. Two schedules are presented: \EE\/ with bandwidth $s=1.5$ (to the left) and an optimal schedule with unit bandwidth (to the right). Time flies downwards. Four lines to the right of each schedule represent each request's lifetime; the bandwidth allotted to each request is outlined in their respective color. \EE\/ first allots all the bandwidth to $S_1$ and splits it evenly among its items $A$, $B$ and $C$ (items $A$, $B$, and $C$ get darker and darker as their broadcasts progress). At time $1$, $S_2$ arrives and \EE\/ splits the bandwidth  &
\raisebox{-8.5cm}{\includegraphics[width=11.5cm]{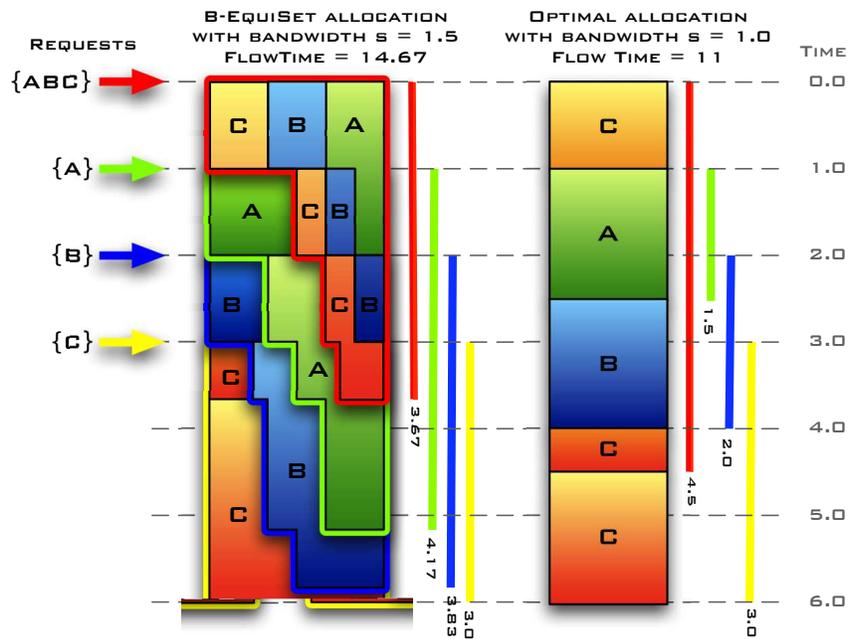}}
\\[-.3em]
\multicolumn{2}{p{\textwidth}}{\footnotesize evenly between  $S_1$ and $S_2$, thus item $A$ is broadcast at a rate: $1.5\times (\frac12 + \frac 12\times \frac13) = 1$ and its broadcast completes at time $2$. At time $2$, $S_3$ arrives, and \EE\/ splits the bandwidth evenly between $S_1$, $S_2$ and $S_3$; $S_1$  has completed its download of $A$, thus \EE\/ splits the bandwidth allotted to $S_1$ among $B$ and $C$ only; $S_2$ was too late to download $A$, so it starts a new broadcast of $A$. $S_1$, $S_2$, $S_3$, and $S_4$ are finally served at time $3+\frac23$, $5+\frac16$, $5+\frac56$ and $6$, for a total flow time $\EE_{1.5}(\bset)=14+\frac23$ whereas~$\bopt_1 = 11$.}
\end{tabular}
\caption{An $1.5$-speed execution of a \EE\/ algorithm.}
\label{fig:ee:1.5}
\end{figure}

\smallskip
Figure~\ref{fig:ee:1.5} illustrates an execution of the algorithm, in which \EE\/ chooses for each alive request $S_j$, to divide up the bandwidth allotted to $S_j$ equally among every $S_j$'s alive items. 

Note that bandwidth adjustments for each item are  necessary only when new requests arrive or when the broadcast of some item completes.

\smallskip

As in \cite{EdmondsPruhs2003}, we deduce the performances of our broadcast algorithm \EE\/ from the analysis of the performances of an other algorithm, \EA\/, for a variant of the non-clairvoyant scheduling problem studied in~\cite{Edmonds1999} which includes dependencies.  Section~\ref{sec:EA} presents this later problem and analyzes the competitiveness of algorithm \EA. Then, Section~\ref{sec:EE}  deduces the competitiveness of \EE\/ by simulating \EA\/ on a particular instance of non-clairvoyant scheduling built on the execution of \EE.

\section{Non-Clairvoyant Seq-Par Batch Scheduling} \label{sec:EA}

For the sake of completeness we first sum up the results in~\cite{Edmonds1999}, reader may skip this paragraph in a first reading. Edmonds's non-clairvoyant scheduling problem consists in designing an online algorithm that schedules jobs on $p$ processors without any knowledge of the progress of each job before its completion. An instance of non-clairvoyant job scheduling problem consists in a collection of jobs $(J_k)$ with arrival times $(a_k)$; each job $J_k$ goes through a series of phases $J_k^1,\ldots, J_k^{m_k}$; the amount of work in each phase $J_k^l$ is $w_k^l$; at time $t$, the algorithm allocates to each uncompleted job $J_k$ an amount $\rho_k^t$ of processors (the $(\rho_k^t)$s are arbitrary non-negative real numbers, such that at any time: $\sum_k \rho_k^t \leq p$); each phase $J_k^l$ progresses at a rate given by a speed-up function $\Gamma_k^l(\rho_k)$ of the amount $\rho_k$ of processors allotted to $J_k$ during phase $J_k^l$, that is to say that the amount of work accomplished between $t$ and $t+dt$ during phase $J_k^l$ is $\Gamma_k^l(\rho_k^t) dt$; let $t_k^l$ denote the completion time of the $l$-th phase of $J_k$, \ie\/ $t_k^l$ is the first time $t'$ such that $\int_{t_k^{l-1}}^{t'} \Gamma_k^l(\rho_k^t)\,dt = w_k^l$ (with $t_k^0= a_k$). The overall goal is to minimize the \emph{flow time} of the jobs, that is to say  the sum of the processing time of each job, \ie\/ $\JFT=\sum_k (t_k^{m_k}-a_k)$. We denote by $\jopt_s(\jset)$ the flow time of an optimal $s$-speed schedule for $\jset$. The algorithm is \emph{non-clairvoyant} in the sense that it does not know anything about the progress of each job and is only informed that a job is completed at the time of its completion. In particular, it is not aware of the different phases that the job goes through (neither of the amount of work nor of the speed-up function). One of the striking results of \cite{Edmonds1999} is that in spite of this total lack of knowledge, the algorithm \EQ\/ that allocates an equal amount of processors to each uncompleted job is $(2+\epsilon)$-speed $(2+4/\epsilon)$-competitive when the speed up functions are arbitrary non-decreasing sub-linear functions (\ie, such that for all $\rho<\rho'$, ${\Gamma_k^l(\rho)}\big/{\rho}\geq {\Gamma_k^l(\rho')}\big/{\rho'}$, for all $k,l$).

Two particular kinds of phases are of interest for our purposes: sequential and parallel. During a \emph{sequential} phase, $\Gamma(\rho)=1$, that is to say that the job progresses at a unit rate whatever amount of processing power it receives (even if it receives no processor at all, \ie\/ even if $\rho = 0$)! During a \emph{parallel} phase, the job progresses proportionally to the processing power it receives, \ie\/ $\Gamma(\rho) = \rho$. Remark that these two kinds of speed-up functions match the requirement of Edmond's theorem and thus \EQ\/ is $(2+\epsilon)$-speed $(2+4/\epsilon)$-competitive on instances consisting of a collection of jobs composed of sequential and parallel phases.

As in \cite{EdmondsPruhs2003}, we reduce the analysis of our broadcast algorithm \EE\/ to the analysis of a non-clairvoyant scheduling algorithm. For that purpose, we need to introduce dependencies between the jobs in Edmonds's framework. We consider the following variant of the \mbox{non-clairvoyant} scheduling problem.\\[-2em] 

\paragraph{Non-Clairvoyant Seq-Par Batches Scheduling.} 
An instance of this variant consists in a collection $\Bset=\{B_1,\ldots,B_q\}$ of batches $B_j=\{J_{j,1},\ldots,J_{j,u_j}\}$ of jobs with arrival times $a_1,\ldots, a_q$, where each job $J_{j,i}$ is composed of two phases: a sequential phase of work $w_{j,i}^s\geq 0$ followed by a parallel phase of work $w_{j,i}^p\geq 0$. (Note that this problem is different from the classical batch scheduling problem in which \emph{only one} batch has to be treated.) The scheduler is non-clairvoyant and discovers each batch of jobs at the time of its arrival and is in particular \emph{not aware of the amounts of work of each job in each batch}.
The scheduler allocates to each job $J_{j,i}$, arrived and uncompleted at time $t$, a certain amount $\rho_{j,i}^t$ of the processors ($\rho_{j,i}^t$ is an arbitrary non-negative real number). Let $t_{j,i}$ denote the completion time of job $J_{j,i}$\,; $t_{j,i}$ is the first date verifying $\int_{a_j+w_{j,i}^s}^{t_{j,i}} \rho_{j,i}^t\, dt=w_{j,i}^p$. We say that a batch is completed as soon as all its jobs are completed; let $t_j$ denote the completion time of batch $B_j$, $t_j = \max_{i=1,\ldots,u_j} t_{j,i}$. The goal is to minimize the \emph{flow time} of the batches, \ie\/ $\BFT=\sum_{B_j\in\Bset} (t_j-a_j)$. We denote by $\Bopt_s(\Bset)$ the flow time of an optimal $s$-speed schedule for $\Bset$.

Similarly to the broadcast setting, we say that a request $B_j$ (resp., a job $J_{j,i}$) is \emph{alive} at time $t$ if $a_j\leq t \leq t_j$ (resp., $a_j \leq t \leq t_{j,i}$).\\[-2em]

\paragraph{\EA\/ Algorithms Family.} Given a \emph{job} scheduling algorithm $A$, we define the \emph{batches} scheduling algorithm \EA\/ as follows. Let $R(t)$ denote the set of batches that are alive at time~$t$. \EA\/ allots to each batch alive at time $t$ an equal amount of processors, \ie, $p/|R(t)|$; then, it runs algorithm $A$ on each alive batch $B_j$ to decide how to split the amount of processors alloted to $B_j$ among its own alive jobs $J_{j,i}$.  In the following, we only require algorithm $A$ to be \emph{fully active}, \ie, that it \emph{allots at all time all the amount of processors} it is given to the alive jobs (\ie, never idles on purpose). Under this requirement, our results hold independently of the choice of $A$. Examples of fully active algorithms $A$ are: $A=\EQ$ which equally splits the amount of processors; or $A=\MINIDX$ which allots all the amount of processors to the smallest indexed alive job $J_{j,i}$ in $B_j$, \ie\/ $i=\min\{i': \text{$J_{j,i'}$ is alive at time $t$}\}$.\\[-2em] 

\paragraph{Analysis of \EA.} 
To analyze the competitiveness of \EA, we associate to each batches scheduling instance $\Bset$, two instances, $\jset'$ and $\jset''$, of job scheduling. We first bound the performances of our algorithm \EA\/ on $\Bset$ from above by the performances of \EQ\/ on $\jset'$ (Lemma~\ref{lem:EA:EQ}). We then use the ``harder'' job instance $\jset''$ to show that the job instance $\jset'$ was in fact ``easier'' than the batch instance $\Bset$ if one increases slightly the number of processors (Lemmas~\ref{lem:JOPT':JOPT''} and~\ref{lem:JOPT'':BOPT}). Since \EQ\/ is competitive on $\jset'$, we can then conclude on the competitiveness of \EA\/ on \Bset\/ (Theorem~\ref{thm:EA}).

Consider a Seq-Par batches scheduling instance $\Bset=\{B_1,\ldots, B_q\}$ where each batch $B_j=\{J_{j,1},\ldots, J_{j,u_j}\}$ arrives at time $a_j$ and each $J_{j,i}$ in $B_j$ consists of a sequential phase of work $w_{j,i}^s$ followed by a parallel phase of work $w_{j,i}^p$. Consider the $s$-speed schedule obtained by running algorithm \EA\/ on instance $\Bset$; let $\rho_{j,i}^t$ denote the amount of processors allotted by \EA\/ to job $J_{j,i}$ at time~$t$, and $\rho_j^t = \sum_{J_{j,i}\in B_j} \rho_{j,i}^t$ denote the amount of processors allotted to batch $B_j$ at time $t$; let $t_{j,i}$ (resp., $t_{j}$) be the completion time of job $J_{j,i}$ (resp., batch $B_j$). We define a Seq-Par job scheduling instance $\jset'=\{J'_1,\ldots,J'_q\}$, where each job $J'_j$ arrives at time~$a_j$, and is composed of a sequential phase of work ${w'_j}^s=\max_{J_{j,i}\in B_j} w_{j,i}^s$, followed by a parallel phase of  work ${w'_j}^p=\int_{a_j+{w'_j}^s}^{t_j} \rho_j^t\, dt$; intuitively, ${w'_j}^s$ is the length of the longest sequential phase among the jobs in $B_j$ and ${w'_j}^p$ is the total amount of parallel work in $B_j$ to be scheduled by \EA\/ after the completion of the last sequential phase among the jobs in $B_j$. 

The key to the next lemma is that one gets exactly the same job schedule of the jobs in $\jset'$ by running algorithm \EQ\/ on instance $\jset'$ as by alloting at all time to each job $J'_j$ the same amount of processors as the jobs in $B_j$ received from \EA\/.

\begin{lemma}[Reduction to job scheduling]
\label{lem:EA:EQ}
%:lem:EA:EQ
If $A$ is fully active, then $\displaystyle \EsA(\Bset) = \EQ_s(\jset')$.
\end{lemma}

\begin{proof}
As long as the longest sequential phase among the jobs in batch $B_j$ is not completed, the batch $B_j$ is alive. By construction, job $J'_j$ is also alive as long as this sequential phase is not completed. Since the amount of processors given to batch $B_j$ in \EA\/ is given by \EQ, and since  \EQ\/ is non-clairvoyant, \EA\/ allots the same amount of processors to $B_j$ as \EQ\/ allots to $J'_j$ until the completion of the longest sequential phase among the jobs in batch $B_j$. By construction, the longest sequential phase in batch $B_j$ and the sequential phase of $J'_j$ end at the same time and at this moment, all the jobs alive in $B_j$ are in their parallel phase. Thus by construction, the overall amount of remaining parallel work in $B_j$ at that time is equal to the parallel work assigned to $J'_j$. By construction, the amount of processors given to $J'_j$ equals the amount of processors alloted to batch $B_j$ which is in turn equal to the total amount alloted to each of its remaining alive jobs since $A$ is fully active. The overall remaining amount of parallel work is thus identical in $J'_j$ and $B_j$ until they complete at the same time. Their flow times are thus identical in both schedules. We conclude the proof by reasoning inductively on the completion times (sorted in non-decreasing order) of each phase of each job in each batch.
\end{proof}

\smallskip

We now define the job instance $\jset''=\{J''_1,\ldots,J''_q\}$. $\jset''$ is a kind of worst case instance of the batch instance $\Bset$, where all the parallel work in each batch $B_j$ has to be scheduled after the longest sequential phase in $B_j$. Job $J''_j$ arrives at time $a_j$ and consists of a sequential phase of work ${w''_j}^s = \max_{J_{j,i}\in B_j} w_{j,i}^s$, followed by a parallel phase of work ${w''_j}^p=\sum_{J_{j,i}\in B_j} w_{j,i}^p$.

\begin{lemma}[$\jset'$ is easier than $\jset''$] \label{lem:JOPT':JOPT''}
$\jopt_s(\jset') \leq \jopt_s(\jset'').$
\end{lemma}

\begin{proof}
Since for all $j$, the sequential works of jobs $J'_j$ and $J''_j$ are identical and  the parallel work in $J'_j$ is bounded from above by the parallel work in $J''_j$, any schedule of $\jset''$ is valid for $\jset'$.
\end{proof}

\begin{lemma}[$\jset''$ with $\delta$ extra processors is ``almost as easy'' as $\Bset$] 
\label{lem:JOPT'':BOPT}
%:lem:JOPT'':BOPT
For all $\delta>0$, $${\jopt_{1+\delta}(\jset'') \leq (1+1/\delta) \Bopt_1(\Bset)}.$$
\end{lemma}

\begin{proof}
The proof consists in showing that when $\delta$ extra processors are given, delaying the completion of each batch $B_j$ by a constant factor, $(1+1/\delta)$, allows to postpone the schedule of all the parallel job phases in  $B_j$ after the completion of the last sequential phase in $B_j$, which concludes the proof by construction of $\jset''$.    

Sort the batches of $\Bset$ by non-increasing arrival time, \ie, assume $a_1 \geq a_2 \geq \ldots \geq a_q$. Consider an optimal schedule $\Bopt_1$ of batches $B_1,\ldots,B_q$ on one processor. We show by induction that there exists a schedule $\jsch$ of $\jset''$ on $1+\delta$ processors such that each job $J''_j$ completes before time $t_j+f_j/\delta$, where $t_j$ and $f_j = t_j - a_j$ denote the completion time and the flow time of $B_j$ in $\Bopt$, respectively. We now show that the parallel phase of each job $J''_j$ can be scheduled between time $t_j$ and $t_j+f_j/\delta$; this concludes the proof since, by construction, the sequential phase of $J''_j$ is necessarily completed before $t_j$.  Start with the first job $J''_1$. Clearly, ${w''_1}^p \leq f_1$. Thus, the total parallel phase of $J''_1$ can be scheduled on the $\delta$ extra processors between time $t_1$ and $t_1+f_1/\delta$. Assume now that the parallel phases of jobs $J''_1,\ldots, J''_{j-1}$ have been scheduled in $\jsch$ during the time intervals $[t_1,t_1+f_1/\delta],\ldots,[t_{j-1},t_{j-1}+f_{j-1}/\delta]$ respectively, and consider job $J''_j$. Since the jobs are considered in non-increasing arrival times, each job $J''_k$ whose parallel phase has been scheduled in $\jsch$ between $t_j$ and $t_j+f_j/\delta$ arrived in the time interval $T=[a_j,t_j+f_j/\delta]$ and furthermore 
$t_k\leq t_j+f_j/\delta$. The total parallel work $W$ of all the jobs currently scheduled in $\jsch$ during $T$, is then in fact scheduled completely in $\Bopt_1$ during $T$. Note that the parallel work of $J''_j$ was also scheduled in $\Bopt_1$ during this time interval. Since $\Bopt_1$ uses only one processor, we conclude that  $W+{w''_j}^p \leq t_j+f_j/\delta-a_j = (1+1/\delta)f_j$. As one can schedule up to $(1+\delta)f_j/\delta = (1+1/\delta)f_j$ parallel work between time $t_j$ and $t_j+f_j/\delta$ on $1+\delta$ processors, the parallel work ${w''_j}^p$ of $J''_j$ can be scheduled in $\jsch$ on time.
\end{proof}

We can now conclude the analysis of \EA.

\begin{theorem}[Competitiveness of \EA] 
\label{thm:EA}
%:thm:EA
For all $\epsilon>0$ and $\delta>0$, $\EA$ is a $(2+\epsilon)(1+\delta)$-speed $(2+4/\epsilon)(1+1/\delta)$-competitive algorithm for the Non Clairvoyant Seq-Par Batches Scheduling problem.
\end{theorem}

\begin{proof}
We use the result of \cite{Edmonds1999} on the competitiveness of \EQ\/ for the non-clairvoyant job scheduling problem to conclude the proof: $
\ESA{(2+\epsilon)(1+\delta)}(\Bset)	
\underset{\makebox[1cm]{\text{\scriptsize(Lemma~\ref{lem:EA:EQ})}}}{=} \EQ_{(2+\epsilon)(1+\delta)}(\jset')
\underset{{\text{\scriptsize(Theorem~1 in \cite{Edmonds1999})}}}{\leq} {(2+4/\epsilon) \, \jopt_{(1+\delta)}(\jset')} 
\underset{\makebox[1cm]{\text{\scriptsize(Lemma~\ref{lem:JOPT':JOPT''})}}}{\leq} (2+4/\epsilon) \, \jopt_{(1+\delta)}(\jset'') 
\underset{\makebox[1cm]{\text{\scriptsize(Lemma~\ref{lem:JOPT'':BOPT})}}}{\leq} (2+4/\epsilon)(1+1/\delta) \, \Bopt_{1}(\Bset)$.
\end{proof}

%%%%%%%%%%%

\section{Competitiveness of \EE} \label{sec:EE}

Consider an instance  of the online data broadcast problem with dependencies: a set $\bset=\{S_1,\ldots, S_q\}$ of $q$ requests with arrival times $a_1,\ldots,a_q$, over $n$ items $I_1,\ldots,I_n$ of lengths $\ell_1,\ldots,\ell_n$. Let $\bsch_s$ be the $s$-speed schedule  designed by \EE\/ on instance $\bset$, and $\EE_s(\bset)$ be its flow time. Let $\osch_1$ be a $1$-speed optimal schedule of $\bset$, and $\bopt_1(\bset)$ be its flow time. 

Following the steps of \cite{EdmondsPruhs2003}, we define an instance $\Bset$ of non-clairvoyant seq-par batches scheduling from $\bsch_s$ and $\osch_1$, such that the performances of \EE\/ on \bset\/ can be compared to the performances of \EA\/ on \Bset\/ for a particular fully-active algorithm $A$. More precisely, we construct \Bset\/ such that 1) the flow time of \EA\/ on \Bset\/ bounds from above the flow time of \EE\/ on \bset\/ and 2) the (batches) optimal flow time for \Bset\/ is at most the (broadcast) optimal flow time for \bset\/ if it is given extra resources. Since \EA\/ is competitive, we can then bound the performances of \EE\/ with respect to the (batches) optimal flow time of \Bset\/ which is by 2) bounded by the (broadcast) optimal flow time of \bset\/.

The intuition behind the construction of $\Bset$ is the following. A batch of all-new jobs is created for each newly arrived request, with one job per requested item. Each job $J$ stays alive until its corresponding item $I$ is served in $\bsch_s$. $J$ is assigned at most two phases depending on the relative service times of $I$ in $\bsch_s$ and $\osch_1$. The sequential phase of $J$ lasts until either $I$ is served in $\bsch_s$, or the broadcast of $I$ starts in $\osch_1$. Intuitively, this means that it is useless to assign processors to $J$ before the optimal schedule does. At the end of its sequential phase, if $J$ is still alive, its parallel phase starts and lasts until the broadcast of $I$ is completed in $\bsch_s$; the parallel work for $J$ is thus defined as the total amount of bandwidth that its corresponding item $I$ received within $J$'s corresponding (broadcast) request in $\EE$. By construction, with a suitable choice of $A$,  \EA\/ constructs the exact same schedule as \EE\/ and claim 1) is verified. Concerning claim 2), the key is to consider the jobs corresponding to the broadcast requests for a given item $I$ that are served by a given broadcast of $I$ in $\osch_1$ starting at some time $t$. The only jobs among them that will receive a parallel phase, are the one for which the broadcast of $I$ in $\bsch_s$ starts just before or just after $t$. By construction, the total amount of parallel work assigned to these jobs corresponds to the bandwidth assigned to the two broadcasts of item $I$ by $\bsch_s$ that start just before and just after time $t$, each of them being bounded by the length of $I$. The total amount of parallel work in the jobs for which the broadcast of the corresponding item $I$ starts in $\osch_1$ at some time $t$, is then bounded by twice the length of $I$, and can thus be scheduled during the broadcast of $I$ in $\osch_1$  if one doubles the number of processors, which proves claim 2).  

The following formalizes the reasoning exposed above.\\[-2em]

\paragraph{The Job Set Instance $\jset$.}
Recall the broadcast instance $\bset$, and the two broadcast schedules $\bsch_s$ and $\osch_1$, defined at the beginning of this section, as well as the notations given in Section~\ref{sec:not}. In particular, let $C^\bsch_s(I_i,t)$ denote the completion time of the broadcast of item $I_i$ that starts just after $t$ in $\bsch_s$, and $B^\osch_1(I_i,t)$ be the time of the beginning of the first broadcast of item $I_i$ that starts after $t$ in  $\osch_1$ (see Section~\ref{sec:not}). Recall the description of algorithm \EE\/ in Section~\ref{sec:alg:EE}: at time $t$, let $R(t)$ be the set of alive requests;  \EE\/ splits equally the bandwidth $s$ among the alive requests and for each alive request $S_j$, it assigns an arbitrary rate $r_{j,i}(t)$ to each alive item $I_i$ in $S_j$, such that $\sum_{\text{$I_i$ alive in $S_j$}} r_{j,i}(t) = s/|R(t)|$; \EE\/ broadcasts then each item $I_i$ at a rate $r_i(t) = \sum_{j} r_{j,i}(t)$ at time $t$.

Given \bset, $\bsch_s$ and $\osch_1$, we define the non-clairvoyant batches scheduling instance $\Bset=\{B_1,\ldots,B_q\}$, where each batch $B_j$ is released at the same time as $S_j$, \ie\/ at time $a_j$, and contains one seq-par job $J_{j,i}$ for each item $I_i\in S_j$ (note that the indices $i$ of the jobs $J_{j,i}$ in each batch $B_j$ may not be consecutive depending on the content of $S_j$). Each job $J_{j,i}$ consists of a sequential phase of work ${w_{j,i}^s = (\min\{ C^\bsch_s(I_i,a_j), B^\osch_1(I_i, a_j)\} - a_j)}$, followed by a parallel phase of work $w_{j,i}^p$. If $C^\bsch_s(I_i,a_j) \leq  B^\osch_1(I_i, a_j)$, then $w_{j,i}^p = 0$; otherwise, $w_{j,i}^p = \int^{C^\bsch_s(I_i,a_j) }_{B^\osch_1(I_i, a_j)} r^{\bsch_s}_{j,i}(t)\, dt + \eta$ where $\eta$ is an infinitely small amount of work, \ie\/ if the download of item $I_i$ in request $S_j$ is completed in $\bsch_s$ after it starts in 
$\osch_1$, then the amount of parallel work assigned to $J_{j,i}$ is just slightly higher than the total amount of bandwidth allotted to item $I_i$ within the bandwidth allotted to request $S_j$ by $\EEs$ after the beginning of the corresponding broadcast in $\osch_1$. Adding an infinitely small amount of work $\eta$ to the parallel phase of $J_{j,i}$ does not change the optimal batches schedule (except on a negligible (discrete) sets of dates) but since the algorithm \EA\/ is non-clairvoyant, this ensures that the job $J_{j,i}$ remains alive until the broadcast of item $I_i$ completes even if $\EEs$ deliberately chooses not to broadcast item $I_i$ in the bandwidth allotted to request $S_j$ (the introduction of infinitely small extra load can be rigorously formalized by adding an exponentially decreasing extra load $\gamma/2^k$ to the $k$th requested job for a small enough~$\gamma$).

\begin{lemma} 
\label{lem:EE:EA}
%:lem:EE:EA
There exists a fully-active algorithm $A$ such that:
$\EE_s(\bset) \leq \EsA(\Bset)$.
\end{lemma}

\begin{proof}
The proof follows the lines of \cite{EdmondsPruhs2003}. Given an amount of processors $\rho$ for an alive batch $B_j$, algorithm $A$ assigns to each alive job $J_{j,i}$ in $B_j$ at time $t$ the same amount of processors as \EEs\/ would have assigned at time $t$ to the 
corresponding alive item $I_i$ of the corresponding alive request $S_j$ which would have been assigned a bandwidth $\rho$.
Since \EEs\/ allots all the bandwidth available to alive jobs, $A$ is fully-active. Now, since $\eta$ is infinitely small, this extra load does not affect the allocation of processors computed by $\EsA$ except over a negligible (discrete) set of dates. By immediate induction, each job $J_{j,i}$ remains alive in the schedule computed by \EsA\/, as long as item $I_i$ is alive in batch $B_j$ in $\bsch_s$. This is clear as long as $J_{j,i}$ is in its sequential phase. Once $J_{j,i}$ enters its parallel phase, as long as the broadcast of item $I_i$ is not completed, either $I_i$ is broadcast by \EEs\/ in batch $B_j$ and  $J_{j,i}$ is scheduled by \EsA\/ ($A$ copies \EEs), or \EEs\/ \emph{deliberately} chooses not to broadcast the alive item $I_i$ and since $J_{j,i}$ has an infinitely small amount of extra work, $J_{j,i}$ remains alive in \EsA\/ as well. The flow time for each job $J_{j,i}$ is then at least the flow time of the corresponding item $I_i$ in $\bsch_s$; we conclude that each batch $B_j$ completes in \EsA\/ no earlier than its corresponding request $S_j$ in \EEs.  
\end{proof}

\begin{lemma}
\label{lem:U2}
%:lem:U2
There exists a $2$-speed batches schedule $\Utwo$ such that:
$\Utwo(\Bset) \leq \FT(\osch_1)$.
\end{lemma}

\begin{proof}
Again, the proof follows the lines of \cite{EdmondsPruhs2003}.  Consider an item $I_i$. We partition the requests $S_j$ containing item $I_i$ into classes $\CC_1,\CC_2,\ldots$, one for each broadcast of $I_i$ in $\osch_1$. The $k$-th class $\CC_k$ contains all the requests $S_j$ that download $I_i$ in $\osch_1$ during its $k$th broadcast, \ie\/ all requests $S_j$ such that $b^\osch_1(I_i,k-1) < a_j \leq b^\osch_1(I_i,k)$ (see Section~\ref{sec:not} for notations).  We show that for all $k$, the total parallel phases of the jobs $J_{j,i}$ such that $S_j\in \CC_k$, can be shoehorned into twice the area of bandwidth allotted by $\osch_1$ to the $k$th broadcast of item $I_i$. Since this holds for all $i$ and all $k$, we obtain a $2$-speed schedule $\Utwo$ such that $\Utwo(\Bset)\leq \FT(\osch_1)$. 

Let $t_1 = b^\osch_1(I_i,k)$ be the time of the beginning of the $k$th broadcast of $I_i$ in $\osch_1$. Consider a request $S_j$ in class $\CC_k$, clearly $a_j\leq t_1$. By construction, job $J_{i,j}$  is assigned a non-zero parallel work only if $S_j$ completes the download of $I_i$ after $t_1$ in $\EEs$. Since $S_j$ arrives before $t_1$, it downloads $I_i$ during one of the two broadcasts of $I_i$ in $\EE_s$ that start just before or just after $t_1$; let $\CC_k^-$ (resp. $\CC_k^+$) be the set of requests served by the broadcast that starts just before $t_1$ (resp. just after $t_1$). Let $t_2$ and $t_3$ be the completion times of the broadcast of $I_i$ in \EEs\/ that start just before and just after $t_1$ respectively. By construction, the total amounts $W^-$ and $W^+$ of parallel work assigned to the jobs $J_{j,i}$ such that $S_j\in\CC_k^-$ and $\CC_k^+$ are respectively: $W^- = \displaystyle \sum_{j\,:\,S_j\in\CC_k^-} \int_{t_1}^{t_2} r_{j,i}(t)\,dt$ and $W^+ =  \displaystyle \sum_{j\,:\,S_j\in\CC_k^+}  \int_{t_1}^{t_3} r_{j,i}(t)\,dt$. Let us rewrite $W^- + W^+ = R_1 + R_2$ with $R_1 =\int_{t_1}^{t_2}  \sum_{j\,:\,S_j\in\CC_k}  r_{j,i}(t)\,dt \leq \int_{t_1}^{t_2} r_i(t)\, dt $ and $R_2 = \int_{t_2}^{t_3} \sum_{j\,:\,S_j\in\CC_k^+} r_{j,i}(t)\,dt \leq \int_{t_2}^{t_3} r_i(t)\, dt$. $R_1$ and $R_2$ are thus at most the total area alloted to item $I_i$ by \EEs\/ during the broadcasts of $I_i$ that start just before and just after $t_1$; since a broadcast is completed as soon as the rates sum up to the length of the items, $R_1\leq \ell_i$ and $R_2\leq \ell_i$, and thus $W^-+W^+ \leq 2\ell_i$. Since $\osch_1$ allots a total bandwidth of $\ell_i$ to broadcast item $I_i$ after time $t_1$,  and since the parallel works of the jobs $J_{j,i}$ such that $S_j\in \CC_k$ are released at time $t_1$ and sum up to a total $W^-+W^+\leq 2\ell_i$, one can construct on 2 processors, a $2$-speed schedule $\Utwo$ in which the parallel phases of each of these jobs $J_{j,i}$  completes before the $k$th broadcast of $I_i$ completes in $\osch_1$. 

Since no processor needs to be allotted to the sequential phases, repeating the construction for each item $I_i$ yields a valid $2$-speed schedule $\Utwo$ in which each job $J_{i,j}$ completes before the corresponding request $S_j$ completes the download of $I_i$ in $\osch_1$. It follows that each batch $B_j$ is completed in $\Utwo$ before its corresponding request $S_j$ is served by $\osch_1$.   
\end{proof}

We now conclude with the proof of the main theorem.

\begin{proofOf}{Theorem~\ref{thm:EE}}
Setting $s=(4+\epsilon)(1+\delta)$, the competitiveness of \EA\/ (Theorem~\ref{thm:EA}) concludes the result:
$
\EE_{(4+\epsilon)(1+\delta)}(\bset) 
\underset{\makebox[1cm]{\text{\scriptsize(Lemma~\ref{lem:EE:EA})}}}{\leq} \ESA{(4+\epsilon)(1+\delta)}(\Bset)
\underset{\makebox{\text{\scriptsize(Theorem~\ref{thm:EA})}}}{\leq} {(2+8/\epsilon)(1+1/\delta) \, \Bopt_2(\Bset)}
\leq {(2+8/\epsilon)(1+1/\delta) \, \Utwo(\Bset)}
\underset{\makebox[1cm]{\text{\scriptsize(Lemma~\ref{lem:U2})}}}{\leq} (2+8/\epsilon)(1+1/\delta) \, \bopt_1(\bset)$.
\end{proofOf}

%%%%%%%%%%%%%%%%%%%%%%%%%%%

\paragraph{The \EEDF\/ algorithm.}
\label{sec:EEDF}
We apply the same method as in \cite{EdmondsPruhs2003}. 
Let $s = (4+\epsilon)(1+\delta)²$ and $c = (2+8/\epsilon)(1+1\delta)²$. \EEDF\/ simulates the $s/(1+\delta)$-speed execution of \EE\/ and at each time $t$ such that the broadcast of an item $I_i$ in \EE\/ is completed, it releases an item $I'_i$ of length $\ell_i$ with a deadline $t+(t-t')/\delta$ where $t'$ is the time of the beginning of the considered broadcast of $I_i$ in \EE\/. Then, \EEDF\/ schedules on a bandwidth $s$ each item $I'_i$ according the earliest-deadline-first policy. With an argument similar to Lemma~\ref{lem:JOPT'':BOPT} or \cite{EdmondsPruhs2003}, one can show that a feasible schedule of the items $I'_i$ exists and thus that earliest-deadline-first constructs it which ensures that \EEDF\/ is $s$-speed $c$-competitive. Since earliest-deadline-first preempts the broadcast of an item only when a new item arrives, \EEDF\/ preempts each broadcast at most once on average.
Note that one can avoid long idle period in \EEDF's schedule by broadcasting an arbitrary item $I_i$ alive in \EE\/ at time $t$ if no item  $I'_i$ is currently alive. \\[-2em]
     
%%%%%%%%%%%%%%%%%%%%%%%%%%%

\paragraph{Concluding remarks.}
Several directions are possible to extend this work. First, \EE\/ does not have precise policy to decide in which order one should broadcast the items within each requested set; deciding on a particular policy may lead to better performances (bandwidth and/or competitive ratio). Second, it might be interesting to design a longest-wait-first greedy algorithm in presence of dependencies; \EE\/ shows that the items should not simply receive bandwidth according to the number of outstanding requested sets for this item (the allotted bandwidth depends also on the number of outstanding items within each outstanding set), it is thus a challenging question to design proper weights to aggregate the current waits of the requested sets including a given item.

\bibliographystyle{plain}
\bibliography{biblio-soda}

\appendix

\section{Omitted proof}

\begin{proofOf}{Fact~\ref{fac:yao}}
We use Yao's principle (see \cite{Yao1977,MotwaniRaghavan1995}) to extend the result to randomized algorithms. 
We consider the following probabilistic distribution of requests set over $n$ items: $1+\sqrt n$ requests arrive at time $t=0$; one request asks for an uniform random subset $S_0$ of size $n-\sqrt{n}$ of the $n$ items; and each of the requests $S_1,\ldots, S_{\sqrt n}$ asks for one random distinct item among the $\sqrt n$ remaining items.  Consider again any deterministic algorithm $A$ with bandwidth~$s$. Since $A$ is deterministic and ignores the dependencies, the schedule designed by $A$ schedule is independent of the random instance. At time $t=n/(2s)$, the broadcast of at least $n/2$ items is not completed. Thus, the probability that request $S_j$, for $j\geq 1$, asks for one of these items is at least $1/2$. Then, the expected number of unsatisfied request at time $t=n/(2s)$ is at least $\sqrt{n}/2$. We conclude that the expected flow time for any deterministic algorithm with bandwidth~$s$ under this distribution of request is at least $n\sqrt n/(4s)$. According to Yao's principle, the worst expected flow time of any randomized algorithm over the collection of all the considered instances is at least $n\sqrt n/(4s)$.   But $\bopt_1 \sim \frac32 n$, which concludes that no randomized algorithm is $s$-speed $c$-competitive, for all $s$ and $c<\sqrt{n}/(6s)$.
\end{proofOf}

\end{document}